\documentclass[fleqn,reqno]{amsart}

\usepackage{graphicx,amscd,amsmath,amssymb,verbatim}
\usepackage[dvips]{hyperref}
\usepackage[TS1,OT1,T1]{fontenc}

\begin{document}

\title[Noninertial symmetry of Hamilton's mechanics]{Noninertial symmetry of Hamilton's mechanics}
\author{Stephen G. Low}
\address{www.stephen-low.net}
\email{Stephen.Low@alumni.utexas.net}
\date{\today}

\maketitle
\begin{abstract}

We present a new derivation of Hamilton's equations that shows that
they have a symmetry group $\mathcal{S}p( 2n) \otimes _{s}\mathcal{H}(
n) $. The group $\mathcal{S}p( 2n) $ is the real noncompact symplectic\ \ group
and $ \mathcal{H}( n) $ is mathematically a Weyl-Heisenberg\ \ group
that is parameterized\ \ by velocity, force\ \ and power where power
is the central element\ \ of the group. The homogeneous\ \ Galilei
group $\mathcal{E}( n) \simeq \mathcal{S}\mathcal{O}( n) \otimes
_{s}\mathcal{A}( n) $, where the special orthogonal group $\mathcal{S}\mathcal{O}(
n) \subset \mathcal{S}p( 2n) $ is parameterized\ \ by rotations\ \ and
the abelian group $\mathcal{A}( n) \subset \mathcal{H}( n) $ is
parameterized\ \ by velocity, is the inertial subgroup. 
\end{abstract}
\maketitle
\section{Symmetry group theorem of Hamilton's equations}

Let $\mathbb{P}=\mathbb{R}^{2n+2}$ be an extended phase space with
coordinates $\{z^{a}\}=\{y^{\alpha },e,t\}$ where $a,b=1,...,2n+2$\ \ and
$\alpha ,\beta =1,...,2n$. The $2n$ $y$-coordinates may also be
written\ \ $\{y^{a}\}=\{p^{i},q^{i}\}$\ \ with $i,j=1,....,n$.\ \ In
these coordinates, there is a symplectic metric that may be written
in the forms 
\begin{equation}
\omega =\zeta _{a,b}d z^{a} d z^{b}={\zeta \mbox{}^{\circ}}_{\alpha
,\beta }d y^{\alpha } d y^{\beta }-d e\wedge d t= \delta _{i,j}d
p^{i}\wedge d q^{j}-d e\wedge d t.%
\label{mo: extended symplectic metric}
\end{equation}

\noindent The $2n +2$ dimensional square matrix of components $\zeta
=[\zeta _{a,b}]$ is given by
\begin{equation}
\zeta =\left( \begin{array}{lll}
 \zeta \mbox{}^{\circ} & 0 & 0 \\
 0 & 0 & -1 \\
 0 & 1 & 0
\end{array}\right) ,\ \ \ \zeta \mbox{}^{\circ}=\left( \begin{array}{ll}
 0 & 1_{n} \\
 -1_{n} & 0
\end{array}\right) ,%
\label{mo: matrix extened symplectic metric}
\end{equation}

\noindent and $1_{n}$ is the unit $n$ dimensional square matrix.\ \ Assume
also that there is a degenerate orthogonal line element 
\begin{equation}
\gamma \mbox{}^{\circ}=d t^{2}={\eta \mbox{}^{\circ}}_{a,b}d z^{a}d
z^{b},%
\label{mo: Newtonian time line element}
\end{equation}

\noindent where the ${\eta \mbox{}^{\circ}}_{a,b}$ are the components
of the $2n +2$ dimensional square matrix that is zero except for
a 1 in the lower right hand corner,
\begin{equation}
\ \ \eta \mbox{}^{\circ}=\left( \begin{array}{lll}
 0 & 0 & 0 \\
 0 & 0 & 0 \\
 0 & 0 & 1
\end{array}\right) .%
\label{mo: degenerate orthogonal metric matrix}
\end{equation}

\noindent As $\mathbb{P}=\mathbb{R}^{2n+2}$, the coordinates and
the form of the symplectic metric (2) and degenerate orthogonal
line element (4) are defined globally.\ \ 
\subsection{{\bfseries Theorem }}

Let $\mathbb{P}$ be extended phase space as defined above with symplectic
metric $\omega $ given in (1) and degenerate orthogonal line element\ \ $\gamma
\mbox{}^{\circ}$ given in (3).\ \ Let $\rho$ be a diffeomorphism
$\rho :\mathbb{P}\rightarrow \mathbb{P}:z\mapsto \tilde{z}=\rho
( z) $ that leaves invariant the symplectic metric, $\omega =\rho
^{*}\omega $ and the degenerate orthogonal line element, $\gamma
\mbox{}^{\circ}={\rho }^{*}\gamma \mbox{}^{\circ}$. Then, 

\noindent A) the connected group of transformations on the cotangent
space leaving the symplectic metric and degenerate orthogonal line
element invariant is 
\begin{equation}
\mathcal{H}\mathcal{S}p( 2n) \simeq \mathcal{S}p( 2n) \otimes _{s}\mathcal{H}(
n) ,%
\label{mo: HSp n}
\end{equation}

\noindent where $\mathcal{H}( n) $ is the Weyl-Heisenberg group
and $\mathcal{S}p( 2n) $ is the real noncompact symplectic group\footnote{The
notation various from author to author, this group is often written
as $\mathcal{S}p( 2n,\mathbb{R}) $.}\cite{Hall}. 

\noindent B) locally the diffeomorphisms $\rho $ must have Jacobian
matrices that are an element of $\mathcal{H}\mathcal{S}p( 2n) $,
\begin{equation}
\left[ \frac{\partial \rho ^{a}( z) }{\partial z^{b}}\right] =\Gamma
( z) \in \mathcal{H}\mathcal{S}p( 2n) \ \ \forall  z\in \mathbb{P},%
\label{mo: jacobian group condition}
\end{equation}

\noindent and consequently have a particular functional form that
satisfy a first order set of differential equations that are Hamilton's
equations \cite{Low7}.
\subsection{{\bfseries Comments}}

In coordinates, the metric and line element\ \ pull back under the
mapping\ \ ${\tilde{z}}^{a}=\rho ^{a}( z) $\ \ is
\begin{gather*}
\omega =\zeta _{a,b}d {\tilde{z}}^{a} d {\tilde{z}}^{b}= \zeta _{a,b}\frac{\partial
\rho ^{a}( z) }{\partial z^{c}}\frac{\partial \rho ^{b}( z) }{\partial
z^{d}}d z^{c} d z^{d}
\\\gamma \mbox{}^{\circ}={\eta \mbox{}^{\circ}}_{a,b}d {\tilde{z}}^{a}
d {\tilde{z}}^{b}= {\eta \mbox{}^{\circ}}_{a,b}\frac{\partial \rho
^{a}( z) }{\partial z^{c}}\frac{\partial \rho ^{b}( z) }{\partial
z^{d}}d z^{c} d z^{d}
\end{gather*}

\noindent and so for the metric and line element to be invariant,
the Jacobian matrices must satisfy
\begin{gather}
\zeta _{c,d}= \zeta _{a,b}\frac{\partial \rho ^{a}( z) }{\partial
z^{c}}\frac{\partial \rho ^{b}( z) }{\partial z^{d}}%
\label{mo: jacobian componets symplectic}
\\{\eta \mbox{}^{\circ}}_{c,d}= {\eta \mbox{}^{\circ}}_{a,b}\frac{\partial
\rho ^{a}( z) }{\partial z^{c}}\frac{\partial \rho ^{b}( z) }{\partial
z^{d}}%
\label{mo: jacobian components orthogonal}
\end{gather}

\noindent The proof that follows first shows that the\ \ matrix
$\Gamma ( z) $ that is defined in (0)\ \ and that satisfies these
equations is an element of $\mathcal{H}\mathcal{S}p( 2n) $ and then
that (0) is Hamilton's equations. 
\subsection{{\bfseries Proof of Part A: Symmetry group is }$\mathcal{H}\mathcal{S}p(
2n) $}

The symplectic metric on extended phase space is invariant under
the symplectic group $\mathcal{S}p( 2n+2) $ and the degenerate orthogonal
line $d t^{2}$ element is invariant under the affine group
\begin{equation}
\mathcal{I}\mathcal{G}\mathcal{L}( 2n+1,\mathbb{R}) \simeq \mathcal{G}\mathcal{L}(
2n+1,\mathbb{R}) \otimes _{s}\mathcal{A}( 2n+1) ,\ \ \ \ \ \ \ \ \ \ \ \ \ \mathcal{A}(
m) \simeq \left(  \mathbb{R}^{m},+\right) .
\end{equation}

We show in this section that the connected group that leaves both
the symplectic metric $\omega $ and the degenerate orthogonal metric
$\gamma \mbox{}^{\circ}$\ \ is
\begin{equation}
\mathcal{H}\mathcal{S}p( 2n) \simeq \mathcal{S}p( 2n+2) \cap \mathcal{I}\mathcal{G}\mathcal{L}(
2n+1,\mathbb{R}) .
\end{equation}

\noindent The symplectic metric $\omega$ given in (0) and degenerate
orthogonal line element\ \ $\gamma \mbox{}^{\circ}$ given in (0)
may be written in matrix notation as
\begin{equation}
\omega =d z^{\mathrm{t}}\zeta  d z,\ \ \ \ d t^{2}=d z^{\mathrm{t}}\eta
\mbox{}^{\circ} d z,
\end{equation}

\noindent Using matrix notation, a transformation of the basis is
$d \tilde{z}=\Gamma  d z$,\ \ $\Gamma \in \mathcal{G}\mathcal{L}(
2n+2,\mathbb{R}) $. It leaves invariant the symplectic metric if
\begin{equation}
\Gamma ^{\mathrm{t}}\zeta  \Gamma =\zeta ,%
\label{mo: symplectic invariance}
\end{equation}

\noindent and the degenerate orthogonal line element is invariant
if
\begin{equation}
\Gamma ^{\mathrm{t}}\eta \mbox{}^{\circ} \Gamma =\eta \mbox{}^{\circ}.%
\label{mo: degenerate orthogonal invariance}
\end{equation}

\noindent Expand the $2n+2$ square matrix $\Gamma $ as 
\begin{equation}
\Gamma =\left( \begin{array}{lll}
 \Sigma  & b & w \\
 {}c & a & r \\
 {}d & g & \epsilon 
\end{array}\right) ,%
\label{mo: candidate matrix}
\end{equation}

\noindent where $\Sigma $ is a $2 n$ dimensional square matrix,
$b,w\in \mathbb{R}^{2n}$ are column vectors, $c,d\in \mathbb{R}^{2n}$
are row vectors and $a,r,g,\epsilon \in \mathbb{R}$.\ \ \ Then expanding
the expression for the invariance of the $\eta \mbox{}^{\circ}$,
(0), \cite{Glimore2}
\begin{equation}
\begin{array}{ll}
 \left( \begin{array}{lll}
 0 & 0 & 0 \\
 0 & 0 & 0 \\
 0 & 0 & 1
\end{array}\right)   & =\left( \begin{array}{lll}
 \Sigma ^{\mathrm{t}} & c^{\mathrm{t}} & d^{\mathrm{t}} \\
 {}b^{\mathrm{t}} & a & g \\
 {}w^{\mathrm{t}} & r & \epsilon 
\end{array}\right) \left( \begin{array}{lll}
 0 & 0 & 0 \\
 0 & 0 & 0 \\
 0 & 0 & 1
\end{array}\right) \left( \begin{array}{lll}
 \Sigma  & b & w \\
 c & a & r \\
 d & g & \epsilon 
\end{array}\right)  \\
   & =\left( \begin{array}{lll}
 d^{\mathrm{t}}d & d^{\mathrm{t}}g & d^{\mathrm{t}}\epsilon  \\
 g d & g^{2} & g \epsilon  \\
 d \epsilon  & g \epsilon  & \epsilon ^{2}
\end{array}\right) .
\end{array}
\end{equation}

\noindent This identity requires $d=g=0$ and $\epsilon =\pm 1$.
Applying this to (0), and computing the determinant 
\[
\mathrm{Det} \Gamma =\mathrm{Det} \left( \begin{array}{lll}
 \Sigma  & b & w \\
 {}c & a & r \\
 {}0 & 0 & \epsilon 
\end{array}\right) =\epsilon  \mathrm{Det} \left( \begin{array}{ll}
 \Sigma  & b \\
 {}c & a
\end{array}\right) \neq 0,
\]

\noindent and so $(\begin{array}{ll}
 \Sigma  & b \\
 {}c & a
\end{array})\in \mathcal{G}\mathcal{L}( 2 n + 1, \mathbb{R}) $ with
$(w,r)\in \mathbb{R}^{2n+1}$. 

A group $\mathcal{G}$ is a semidirect product if it has a subgroup
$\mathcal{K}\subset \mathcal{G}$ and a normal subgroup\ \ $\mathcal{N}\subset
\mathcal{G}$ such that $\mathcal{G}\simeq \mathcal{N} \mathcal{K}$
and $\mathcal{K}\cap \mathcal{N}=\text{\boldmath $e$}$\ \ where
$\text{\boldmath $e$}$ is the trivial group. It is straightforward
to verify that the above matrices define the extended affine group
\begin{equation}
\hat{\mathcal{I}\mathcal{G}\mathcal{L}}( 2 n+1, \mathbb{R}) \simeq
\mathbb{Z}_{2}\otimes _{s}\mathcal{I}\mathcal{G}\mathcal{L}( 2 n+1,
\mathbb{R}) , 
\end{equation}

\noindent where the affine group is 
\begin{equation}
\mathcal{I}\mathcal{G}\mathcal{L}( 2 n+1, \mathbb{R}) \simeq \mathcal{G}\mathcal{L}(
2 n+1, \mathbb{R}) \otimes _{s}\mathcal{A}( n+1) .
\end{equation}

The $\mathbb{Z}_{2}$ group, parameterized by $\epsilon =\pm 1$\ \ is
the discrete group that changes the sign of $t$. The affine group
is the maximal connected subgroup. As we only require the connected
component, we can set $\epsilon =1$.

 Next, the symplectic invariance condition (0) requires that
\begin{equation}
\begin{array}{ll}
 \left( \begin{array}{lll}
 \zeta \mbox{}^{\circ} & 0 & 0 \\
 0 & 0 & -1 \\
 0 & 1 & 0
\end{array}\right)   & =\left( \begin{array}{lll}
 \Sigma ^{\mathrm{t}} & c^{\mathrm{t}} & 0 \\
 {}b^{\mathrm{t}} & a & 0 \\
 {}w^{\mathrm{t}} & r & 1
\end{array}\right) \left( \begin{array}{lll}
 \zeta \mbox{}^{\circ} & 0 & 0 \\
 0 & 0 & -1 \\
 0 & 1 & 0
\end{array}\right)  \left( \begin{array}{lll}
 \Sigma  & b & w \\
 c & a & r \\
 0 & 0 & 1
\end{array}\right)  \\
   & =\left( \begin{array}{lll}
 \Sigma ^{\mathrm{t}}\zeta \mbox{}^{\circ}\Sigma  & \Sigma ^{\mathrm{t}}\zeta
\mbox{}^{\circ} b & -c^{\mathrm{t}}+\Sigma ^{\mathrm{t}}\zeta \mbox{}^{\circ}
w \\
 b^{\mathrm{t}}\zeta \mbox{}^{\circ}\Sigma  & 0 & -a +b^{\mathrm{t}}\zeta
\mbox{}^{\circ} w \\
 c+w^{\mathrm{t}}\zeta \mbox{}^{\circ}\Sigma  & a+w^{\mathrm{t}}\zeta
\mbox{}^{\circ} b & 0
\end{array}\right) .
\end{array}
\end{equation}

\noindent This identity is satisfied with
\begin{equation}
b=0, a=1, c= - w^{\mathrm{t}}\zeta \mbox{}^{\circ} \Sigma ,\ \ \Sigma
^{\mathrm{t}} \zeta \mbox{}^{\circ} \Sigma =\zeta \mbox{}^{\circ}.
\end{equation}

\noindent $\Gamma $ now has the form
\begin{equation}
\Gamma ( \Sigma ,w,r) =\left( \begin{array}{lll}
 \Sigma  & 0 & w \\
 {}-w^{\mathrm{t}}\zeta \mbox{}^{\circ} A & 1 & r \\
 {}0 & 0 & 1
\end{array}\right) .%
\label{mo: HSp matrix group}
\end{equation}

\noindent where $\Sigma \in \mathcal{S}p( 2n) $, $w\in \mathbb{R}^{2n}$
and $r\in \mathbb{R}$. 

The group multiplication of the matrix group given by (0) is determined
by matrix multiplication to be
\begin{equation}
\Gamma ( \Sigma ^{{\prime\prime}},w^{{\prime\prime}},r^{{\prime\prime}})
=\Gamma ( \Sigma ^{\prime },w^{\prime },r^{\prime }) \Gamma ( \Sigma
,w,r) ,%
\label{mo: HSP group law}
\end{equation}

\noindent where
\begin{equation}
\begin{array}{l}
 \Sigma ^{{\prime\prime}}= \Sigma ^{\prime }\Sigma , \\
 w^{{\prime\prime}}=w^{\prime }+\Sigma ^{\prime } w, \\
 r^{{\prime\prime}}= r^{\prime }+r-{}{w^{\prime }}^{\mathrm{t}}\zeta
\mbox{}^{\circ} \Sigma ^{\prime } w.
\end{array}%
\label{mo: HSP group law components}
\end{equation}

\noindent and the inverse is determined by the matrix inverse to
be
\begin{equation}
\Gamma ^{-1}( \Sigma ,w,r) =\Gamma ( \Sigma ^{-1},-\Sigma ^{-1}w,-r)
.
\end{equation}

The following groups are subgroups
\begin{equation}
\begin{array}{l}
 \Gamma ( \Sigma ,0,0) \in \mathcal{S}p( 2n) , \\
 \Gamma ( 1_{2n},w,r) =\Upsilon ( w,r) \text{}\in \mathcal{H}( n)
\simeq \mathcal{A}( n) \otimes _{s}\mathcal{A}( n+1) .
\end{array}%
\label{mo: HSp subgroups}
\end{equation}

\noindent where $\mathcal{A}( m) $ is the real abelian group under
addition, $\mathcal{A}( m) \simeq (\mathbb{R}^{m},+)$. It is then
be shown that $\Upsilon ( w,r) \text{}\in \mathcal{H}( n) $\ \ \cite{folland}
is a normal subgroup by computing the automorphisms\ \ \ 
\begin{equation}
\begin{array}{ll}
 \Upsilon ( w^{{\prime\prime}},r^{{\prime\prime}})  & =\begin{array}{ll}
 \Gamma ( \Sigma ^{\prime },w^{\prime },r^{\prime }) \Upsilon (
w,r)  &  \Gamma ^{-1}( \Sigma ^{\prime },w^{\prime },r^{\prime })
\end{array} \\
  & =\Upsilon ( \Sigma ^{\prime } w,r+ {\left( \Sigma ^{\prime }w\right)
}^{\mathrm{t}} \zeta \mbox{}^{\circ} w^{\prime }-{w^{\prime }}^{\mathrm{t}}
\zeta \mbox{}^{\circ} \Sigma ^{\prime } w) .
\end{array}%
\label{mo: HSp Automorphisms of the Heisenberg Group}
\end{equation}

\noindent As 
\begin{equation}
\frac{\Gamma ( 1_{2n},w,r) \cap \Gamma ( \Sigma ,0,0) =\Gamma (
1_{2n},0,0) ,}{\Gamma ( \Sigma ,w,r) =\Gamma ( 1_{2n},w,r) \Gamma
( \Sigma ,0,0) ,}
\end{equation}

\noindent it follows that the intersection of the groups is the
identity and $\mathcal{H}\mathcal{S}p( 2n) \simeq \mathcal{H}( n)
\mathcal{S}p( 2n) $ group is the semidirect product (0) as claimed.

It is straightfoward to show with $\epsilon =\pm 1$ that the intersection
of the symplectic and\ \ extended affine group is
\begin{equation}
\mathcal{S}p( 2n+2) \cap \hat{\mathcal{I}\mathcal{G}\mathcal{L}}(
2 n+1, \mathbb{R}) \simeq \hat{\mathcal{H}\mathcal{S}p}( 2n) ,\ \ \ \hat{\mathcal{H}\mathcal{S}p}(
2n) \simeq \mathbb{Z}_{2}\otimes _{s}\mathcal{H}\mathcal{S}p( 2n)
\end{equation}

\noindent where again the $\mathbb{Z}_{2}$ changes the sign of $t$.

That $\mathcal{H}( n) $ is the Weyl-Heisenberg group may be determined
by computing its algebra
\begin{equation}
W_{a}=\frac{\partial }{\partial  w^{a}}\Upsilon ( w,r) |_{w=r=0},
U=\frac{\partial }{\partial  r}\Upsilon ( w,r) |_{w=r=0}.
\end{equation}

\noindent A general element of the algebra is\ \ $Z=w^{\alpha }W_{\alpha
}+r U$. The Lie algebra of a matrix group is the matrix commutators
$[A,B]=A B-B A$ that give
\begin{equation}
\left[ W_{\alpha },W_{\beta }\right] =2{\zeta \mbox{}^{\circ}}_{\alpha
,\beta }U,\ \ \ \left[ W_{\alpha },U\right] =0.
\end{equation}

\noindent This is the Weyl-Heisenberg algebra where $U$ is the central
generator.\ \ The factor of 2 is just normalization. It can be removed
simpy by scaling $r\mapsto 2 r$.

This completes the proof of Part A of the theorem that establishes
that the connected group that has both\ \ symplectic and affine
symmetry is $\mathcal{H}\mathcal{S}p( 2n) $.
\subsection{{\bfseries Proof of Part B:\ \ Diffeomorphisms\ \ satisfy
Hamilton's\ \ equations}}

The Jacobian matrix $[\frac{\partial \rho ( z) }{\partial z}]$ of
the\ \ diffeomorphism $\rho $ that leaves invariant the symplectic
metric\ \ (0) and the degenerate orthogonal line element (0) must
satisfy (0) and (0). Therefore, the Jacobian matrix is an element
of the symmetry group, $[\frac{\partial \rho ( z) }{\partial z}]=\Gamma
( z) \in \mathcal{H}\mathcal{S}p( n) $.\ \ Using the expanded notation
with $\{z\}=\{y,e,t\}\text{}$with $z\in \mathbb{R}^{2n+2}$, $y\in
\mathbb{P}\mbox{}^{\circ}\simeq \mathbb{R}^{2n}$ and $e,t\in \mathbb{R}$,
and likewise for the diffeomorphism 
\begin{equation}
\left\{ \rho ( z) \right\}  =\left\{ \rho _{y}( y,e,t) ,\rho _{e}(
y,e,t) ,\rho _{t}( y,e,t) \right\} 
\end{equation}

\noindent the Jacobian matrix is
\begin{equation}
\left( \begin{array}{lll}
 \frac{\partial \rho _{y}( y,e,t) }{\partial y} & \frac{\partial
\rho _{y}( y,e,t) }{\partial e} & \frac{\partial \rho _{y}( y,e,t)
}{\partial t} \\
 \frac{\partial \rho _{e}( y,e,t) }{\partial y} & \frac{\partial
\rho _{e}( y,e,t) }{\partial e} & \frac{\partial \rho _{e}( y,e,t)
}{\partial t} \\
 \frac{\partial \rho _{t}( y,e,t) }{\partial y} & \frac{\partial
\rho _{t}( y,e,t) }{\partial e} & \frac{\partial \rho _{t}( y,e,t)
}{\partial t}
\end{array}\right) =\left( \begin{array}{lll}
 \Sigma  \left( z\right)  & 0 &  w( z)  \\
 {}-w^{\mathrm{t}}( z) \zeta \mbox{}^{\circ}\ \ \Sigma  \left( z\right)
& 1 & r( z)  \\
 {}0 & 0 & 1
\end{array}\right) .%
\label{mo: extended jacobi matrix}
\end{equation}

\noindent where we are suppressing indices and using matrix notation.

This restricts the functional dependency of the diffeomorphisms
as follows. First the time component, $\frac{\partial \rho _{t}(
y,e,t) }{\partial y}=\frac{\partial \rho _{t}( y,e,t) }{\partial
e}=0$ and $\frac{\partial \rho _{t}( y,e,t) }{\partial t}=1$ and
so ignoring trivial integration constants, $\rho _{t}( y,e,t) =t.$
Next for the energy component, note that\ \ $\frac{\partial \rho
_{e}( y,e,t) }{\partial e}=1$ and therefore $\rho _{e}$ may be written
as $\rho _{e}( y,e,t) =e+H( y,t) $ where $H$ is some function.\ \ \ Finally,
$\frac{\partial \rho _{y}( y,e,t) }{\partial e}=0$ and consequently
$\rho _{y}( y,e,t) =\varphi ( y,t) $ where\ \ $\varphi $ is some
function.

Summarizing, the diffeomorphism $\tilde{z}=\rho ( z) \text{}$can
be expanded as 
\begin{equation}
\begin{array}{l}
 \tilde{y}=\rho _{y}( y,e,t) =\varphi ( y,t) = \phi _{y}( t) , \\
 \tilde{e}=\rho _{e}( y,e,t) =e+H( y,t) , \\
 \tilde{t}=\rho _{t}( y,e,t) = t.
\end{array}%
\label{mo: functional form of diffeomorphism}
\end{equation}

\noindent $H$ and $\varphi $ are\ \ functions 
\begin{equation}
\begin{array}{l}
 H: \mathbb{R}^{2n+1}\rightarrow \mathbb{R}:\left( y,t\right) \mapsto
H( y,t) , \\
 \varphi : \mathbb{R}^{2n+1}\rightarrow \mathbb{P}\mbox{}^{\circ}:\left(
y,t\right) \mapsto \varphi ( y,t) .
\end{array}
\end{equation}

\noindent $\phi _{y}$ are the curves defined by 
\begin{equation}
\phi _{y}: \mathbb{R}\rightarrow \mathbb{P}\mbox{}^{\circ}:t\mapsto
\phi _{y}( t) =\varphi ( y,t) ,\ \ \phi _{y}( 0) =\varphi ( y,0)
=y.%
\label{mo: phi curves}
\end{equation}

\noindent $H$ will turn out to be the Hamiltonian and $\phi _{y}$
the curves that are the trajectories in phase space that are solutions
to Hamilton's equations.

Substituting these back into (0), the Jacobian now has the form
\begin{equation}
\left( \begin{array}{lll}
 \frac{\partial \varphi ( y,t) }{\partial y} & 0 & \frac{\partial
\varphi ( y,t) }{\partial t} \\
 \frac{\partial H( y,t) }{\partial y} & 1 & \frac{\partial H( y,t)
}{\partial t} \\
 0 & 0 & 1
\end{array}\right) =\left( \begin{array}{lll}
 \Sigma  \left( y,t\right)  & 0 &  w( y,t)  \\
 {}-w^{\mathrm{t}}( y,t) \zeta \mbox{}^{\circ}\ \ \Sigma  \left(
y,t\right)   & 1 & r( y,t)  \\
 {}0 & 0 & 1
\end{array}\right) .%
\label{mo: extended jacobi matrix}
\end{equation}

\noindent Therefore we have
\begin{equation}
\frac{\partial \varphi ( y,t) }{\partial y}=\Sigma  \left( y,t\right)
,\ \ \frac{\partial H( y,t) }{\partial y}=-{
\text{}[ \frac{\partial \varphi ( y,t) }{\partial t}] }^{\mathrm{t}}
\zeta \mbox{}^{\circ}\ \ \Sigma  \left( y,t\right)  ,\ \ \ \frac{\partial
H( y,t) }{\partial t}=r( y,t) 
\end{equation}

\noindent As $\varphi ( y,t) $ is a canonical transformation for
some $y \mbox{}^{\circ}$,\ \ \ $y=\varphi ( y \mbox{}^{\circ},t)
$ and for some $t \mbox{}^{\circ}\text{}$,\ \ $y \mbox{}^{\circ}=\phi
_{y \mbox{}^{\circ}}( t \mbox{}^{\circ}) $ with $\Sigma  (y \mbox{}^{\circ},t
\mbox{}^{\circ})=1_{2n}$.\ \ Then from the chain rule,\ \ 
\begin{equation}
\frac{\partial \varphi ( y,t) }{\partial t}=\frac{\partial \varphi
( y,t) }{\partial y} \frac{\partial \varphi ( y \mbox{}^{\circ},t)
}{\partial t}=\Sigma  \left( y,t\right) \frac{\partial \varphi (
y \mbox{}^{\circ},t) }{\partial t}=\Sigma  \left( y,t\right) \frac{d
\phi _{y \mbox{}^{\circ}}( t) }{d t}
\end{equation}

\noindent Consequently
\begin{equation}
\frac{\partial H( y,t) }{\partial y}=-{
\text{}[ \frac{d \phi _{y \mbox{}^{\circ}}( t) }{d t}] }^{\mathrm{t}}\Sigma
^{\mathrm{t}} \left( y,t\right)  \zeta \mbox{}^{\circ}\ \ \Sigma
\left( y,t\right)  =-{
\text{}[ \frac{d \phi _{y \mbox{}^{\circ}}( t) }{d t}] }^{\mathrm{t}}\zeta
\mbox{}^{\circ}
\end{equation}

\noindent Re-arranging
\begin{equation}
\text{}\frac{d \phi _{y \mbox{}^{\circ}}( t) }{d t}=-{\zeta \mbox{}^{\circ}[
\frac{\partial H( y,t) }{\partial y}] }^{\mathrm{t}},\ \ \ \frac{\partial
H( y,t) }{\partial t}=r( y,t) 
\end{equation}

In component form this is 
\begin{equation}
\text{}\frac{d \phi _{y \mbox{}^{\circ}}^{\alpha }( t) }{d t}={\zeta
\mbox{}^{\circ}}^{\alpha ,\beta }\frac{\partial H( y,t) }{\partial
y^{\beta }},\ \ \ \frac{\partial H( y,t) }{\partial t}=r( y,t) %
\label{mo: Hamilton's equations}
\end{equation}

\noindent where $[{\zeta \mbox{}^{\circ}}^{\alpha ,\beta }]=-\zeta
\mbox{}^{\circ}$.\ \ These are Hamilton's equations with the initial
point $y \mbox{}^{\circ}=\phi _{y \mbox{}^{\circ}}( t \mbox{}^{\circ})
$. 

The converse requires us to prove that if the diffeomorphisms satisfy
Hamilton's equations (0), then the symplectic and line element are
invariant.\ \ \ 
\begin{equation}
\begin{array}{ll}
 \tilde{\omega } & =d {\tilde{y}}^{\mathrm{t}}\zeta \mbox{}^{\circ}
d \tilde{y} +d\tilde{t}\wedge d \tilde{e} \\
   & ={\left( d y+d \phi _{y \mbox{}^{\circ}}( t) \right) }^{\mathrm{t}}\zeta
\mbox{}^{\circ} \left( d y+d \phi _{y \mbox{}^{\circ}}( t) \right)
+ d t\wedge \left( d e+d H( y,t) \right)  \\
   & =d y^{\mathrm{t}}\zeta \mbox{}^{\circ} d y +d t\wedge d e-{\left[
\frac{d \phi _{y \mbox{}^{\circ}}( t) }{d t}\right] }^{\mathrm{t}}
\zeta \mbox{}^{\circ} d y\wedge d t-\ \ \frac{\partial  H( y,t)
}{d y}d y\wedge d t \\
   & =\omega - \left( {\left[ \zeta \mbox{}^{\circ}\frac{d \phi
_{y \mbox{}^{\circ}}( t) }{d t}\right] }^{\mathrm{t}} -\frac{\partial
H( y,t) }{d y}\right) d y\wedge d t \\
   & =\omega 
\end{array}
\end{equation}
$\gamma \mbox{}^{\circ}=d t^{2}$ is invariant as $t$ is an invariant
parameter in Hamilton's equations.\ \ This completes the proof of\ \ the
theorem.

A corollary of the theorem is that Hamilton's equations are valid
in any extended canonical coordinates where the symplectic metric
and degenerate line element have the form given in (0) and (0).
Furthermore, transformations between these extended canonical coordinates
must have a Jacobian that is an element of the $\mathcal{H}\mathcal{S}p(
2n) $ group (0).
\section{Physical meaning of the theorem}

The symplectic symmetry and affine symmetries are very well know
to be fundamental symmetries of classical mechanics.\ \ It should
not therefore be a surprise that the intersection of these symmetries,
where both are manifest, plays a fundamental role in Hamilton's
mechanics. 

An element $\Gamma \in \mathcal{H}\mathcal{S}p( 2n) \simeq \mathcal{S}p(
2n) \otimes _{s}\mathcal{H}( n) $, due to the defining properties
of the semidirect product can always be written as the product of
a symplectic transformation and a Weyl-Heisenberg transformation
\begin{equation}
\Gamma ( \Sigma ,y,r) =\Gamma ( 1_{n},y,r) \Gamma ( \Sigma ,0,0)
.
\end{equation}

We will consider the symplectic group first and show that this is
the standard canonical transforms on phase space.\ \ Next, we consider
the Weyl-Heisenberg transformations and\ \ show that they lead to
familiar results. 
\subsection{Symplectic\ \ transformations}

Consider first the symplectic transformations.\ \ In this case,\ \ the
general transformations (0) reduce to
\begin{equation}
\tilde{y}=\rho _{y}( y,t) =\varphi ( y,t) ,\ \ \tilde{e}=\rho _{e}(
e) =e , \tilde{t}=\rho _{t}( t) =t,
\end{equation}

\noindent with Jacobian satisfying 
\begin{equation}
d \tilde{y} = \frac{\partial  \varphi ( y,t) }{\partial  y} d y=\Sigma
( y,t)  d y .
\end{equation}

The $\varphi ( y,t) $ are time dependent canonical transformations
that appear in all the standard treatments of Hamilton's mechanics.\ \ They
may be regarded as the canonical transformations parameterized by
time on the momentum, position phase space $y\in \mathbb{P}\mbox{}^{\circ}\simeq
\mathbb{R}^{2n}$
\begin{equation}
\varphi _{t}: \mathbb{P}\mbox{}^{\circ}\rightarrow  \mathbb{P}\mbox{}^{\circ}:y\mapsto
\tilde{y}=\varphi _{t}( y) ,
\end{equation}
or as the curves $\phi _{y}:\mathbb{R}\rightarrow \mathbb{P}\mbox{}^{\circ}$
that are given in (0).\ \ The solutions $\phi _{y}$ to Hamilton's
equations may be regarded as a time evolving canonical transformation.

\ \ The coordinates in which the symplectic metric have the canonical
form (0) are canonical coordinates.\ \ In particular, Hamilton's
equations are valid in any canonical coordinates $\tilde{y} =\varrho
( y) $ with
\begin{equation}
d \tilde{y} = \frac{\partial  \varrho ( y) }{\partial  y} d y=\Sigma
( y)  d y .
\end{equation}

\noindent Hamilton's equations in the tilde coordinates are
\begin{equation}
\text{}\frac{d {\tilde{\phi }}_{y \mbox{}^{\circ}}( t) }{d t}=-{\zeta
\mbox{}^{\circ}[ \frac{\partial \tilde{H}( \tilde{y},t) }{\partial
\tilde{y}}] }^{\mathrm{t}},\ \ \ %
\label{mo: Hamilton's tilda equations}
\end{equation}

\noindent with 
\begin{equation}
\tilde{H}( \tilde{y},t) =\tilde{H}( \varrho ( y) ,t) = H( y,t) ,\ \ \ {\tilde{\phi
}}_{y \mbox{}^{\circ}}( t) = \varrho ( \phi _{y \mbox{}^{\circ}}(
t) ) ,
\end{equation}

\noindent and therefore
\begin{equation}
 \tilde{H}=H\circ \varrho ^{-1} \mathrm{and} {\tilde{\phi }}_{y
\mbox{}^{\circ}}=\varrho \circ \phi _{y \mbox{}^{\circ}}.\ \ %
\label{mo: Hamiltonian transformation}
\end{equation}

\noindent It then follows from the methods used to prove the general
theorem that Hamilton's equations transform into the non-tilde coordinates
for the transforms $\varrho $ that are the time independent special
case of the more general $\rho $ transforms of the theorem.

Note particularly that under a canonical transformation, that the
Hamiltonian transforms as\ \ $\tilde{H}=H\circ \varrho ^{-1}$ given
in (0) and not as an invariant function $\tilde{H}=H$.\ \ \ \ Canonical
coordinates do not have the concept of states being inertial or
noninertial and Hamilton's equations are valid in either provided
that the Hamiltonian $H( y,t) $ is chosen appropriately according
to (0).\ \ 

The phase space $\mathbb{P}\mbox{}^{\circ}$ may be generalized to
symplectic manifolds with Hamilton's equations expressed as the
flows of Hamiltonian vector fields \cite{Arnold}. 
\subsection{Weyl-Heisenberg\ \ transformations}

Define $y=(p,q)$, $p,q\in \mathbb{R}^{n}$\ \ and $\phi =(\pi ,\xi
) $, In components, this is $\{y^{a}\}=\{p^{i},q^{i}\}$, $\{\phi
^{a}( t) \}=\{\pi ^{i}( t) ,\xi ^{i}( t) \}$ $i,j=1,..,n$. As is
usual, $p$ is canonical momentum and $q$ is canonical position.
We will continue to use matrix notation with indices suppressed.\ \ Hamilton's
equations then take on their most simple form,
\begin{equation}
\text{}\frac{d \xi ( t) }{d t}=v=\frac{\partial H( p,q,t) }{\partial
p},\text{}\frac{d \pi ( t) }{d t}=f=-\frac{\partial H( p,q,t) }{\partial
q},\ \ \ \frac{\partial H( p,q,t) }{\partial t}=r,%
\label{mo: p q Hamilton's equations}
\end{equation}

\noindent where $v( p,q,t) ,f( p,q,t) \in \mathbb{R}^{n}$ are the
velocity and force respectively and $r( p,q,t) \in \mathbb{R}$ is
the power.\ \ The velocity force and power are generally functions
of $(p,q,t)$ and this will be implicit in the following. The Weyl-Heisenberg
subgroup may be written as 
\begin{equation}
\Upsilon ( f,v,r) =\Gamma ( 1_{2n},f,v,r) =\left( \begin{array}{llll}
 1_{n} & 0 & 0 & f \\
 0 & 1_{n} & 1 & v \\
 v & -f & 1 & r \\
 0 & 0 & 0 & 1
\end{array}\right) ,%
\label{mo: Heisenberg matrix}
\end{equation}

The coordinates $z$ of the extended phase space $\mathbb{P}$ may
be similarly expanded as $z= (p,q,e,t)$ and the Weyl-Heisenberg
transformation $d \tilde{z}=\Upsilon  d z $expands as 
\begin{equation}
\left( \begin{array}{l}
 d \tilde{p} \\
 d \tilde{q} \\
 d \tilde{e} \\
 d \tilde{t}
\end{array}\right) =\left( \begin{array}{llll}
 1_{n} & 0 & 0 & f \\
 0 & 1_{n} & 1 & v \\
 v & -f & 1 & r \\
 0 & 0 & 0 & 1
\end{array}\right) \left( \begin{array}{l}
 d p \\
 d q \\
 d e \\
 d t
\end{array}\right) .%
\label{mo: p q Heisenberg matrix}
\end{equation}

\noindent Using Hamilton's equations (0), this results in
\begin{equation}
\begin{array}{ll}
 d\tilde{t}=d t, &   \\
 d\tilde{q}=d q+v d t & \mathit{=} d q+d \mathrm{\xi }\left( t\right)
, \\
 d\tilde{p}=d p+f d t & \mathit{=} d p+d \mathrm{\pi }\left( t\right)
, \\
 d \tilde{e} = d e+v \cdot d p-f\cdot d q+r d t & = d e+d H( p,q,t)
\mathit{.}
\end{array}%
\label{mo: nonrelativistic noinertial}
\end{equation}

These are the transformations that relate two states in extended
phase space that have a relative rate of change of position, momentum
and energy with respect to time. That is, they have a relative velocity
$v$, force $f$ and power $r$.\ \ These are general states in the
extended phase space that may be inertial or noninertial.\ \ In
the energy transformation, $\int v\cdot d p$ is the incremental
kinetic energy and $-\int f\cdot d q$ is the work transforming from
energy state $e$ to $ \tilde{e} $. The term $\int r d t$ is the
explicit power for time dependent Hamiltonians.\ \ Solving Hamilton's
equations enables these to be integrated to the form that is a special
case of (0) with $\Sigma =1_{2n}$,
\begin{equation}
\begin{array}{l}
 \tilde{t}=\rho _{t}( t)  =t, \\
 \tilde{q}=\rho _{q}( q,t) =q + \xi ( t) , \\
 \tilde{p}=\rho _{p}( p,t) =p + \pi ( t) ,  \\
 \tilde{e} =\rho _{e}( e,p,q,t) = e + H( p,q,t) .
\end{array}%
\label{mo: nonrelativistic noinertial}
\end{equation}

Using the group multiplication (0-0) with $\Sigma =1_{2n}$ , or
simply multiplying the matrices in (0) together shows that\ \ 
\begin{gather}
\Upsilon ( \tilde{f},\tilde{v},\tilde{r}) \Upsilon ( f,v,r) =\Upsilon
( f+\tilde{f},v+\tilde{v},r+\tilde{f} v-\tilde{v} f) ,
\\\Upsilon ( f,v,r) \Upsilon ( \tilde{f},\tilde{v},\tilde{r}) =\Upsilon
( f+\tilde{f},v+\tilde{v},r-\tilde{f} v+\tilde{v} f) .
\end{gather}

These are not equal and consequently the operations do not commute.\ \ This
can be made even more explicit by considering the case of a transformation
in\ \ velocity followed by a transformation in force 
\begin{gather}
\Upsilon ( \tilde{f},0,0) \Upsilon ( 0,v,0) =\Upsilon ( \tilde{f},v,\tilde{f}
v) 
\\\Upsilon ( 0,v,0) \Upsilon ( \tilde{f},0,0) =\Upsilon ( \tilde{f},v,-\tilde{f}
v) 
\end{gather}

This is not unexpected. We do not expect an inertial transformation
in velocity followed by a noninertial transformation in force to
be the same as the noninertial force transformation followed by
the inertial velocity transformation. What is unexpected is that
the noncommutivity is given precisely by the Weyl-Heisenberg nonabelian
group. The noncommutativity is also why noninertial states and frames
are difficult to work with. 
\section{Discussion}

Hamilton's mechanics is a reformulation of Newton's mechanics and
is therefore invariant under Galilean relativity. The homogeneous
Galilei relativity group is mathematically the Euclidean group $\mathcal{E}(
n) \simeq \mathcal{S}\mathcal{O}( n) \otimes _{s}\mathcal{A}( n)
$ parameterized by rotations and velocity. This is a subgroup of
the group of transformations $\mathcal{H}\mathcal{S}p( 2n) $.\ \ The
orthogonal group $\mathcal{S}\mathcal{O}( n) \subset \mathcal{S}p(
2n) $\ \ where in this case the symplectic transformations on $\mathbb{P}\mbox{}^{\circ}$
are just the rotations 
\begin{equation}
 \Sigma ( R) =\left( \begin{array}{ll}
 R & 0 \\
 0 & R
\end{array}\right) .\ \ 
\end{equation}

\noindent The space time translations are a subgroup of the Weyl-Heisenberg
group,\ \ \ $\mathcal{A}( n) \subset \mathcal{H}( n) \simeq \mathcal{A}(
n) \otimes _{s}\mathcal{A}( n+1) $. The resulting transformations
are the inertial transformations on extended phase space 
\begin{equation}
\begin{array}{l}
 d\tilde{t}=d t, \\
 d\tilde{q}=R d q + v d t, \\
 d\tilde{p}=R d p ,  \\
 d \tilde{e} = d e + v \cdot d p. 
\end{array}%
\label{mo: nonrelativistic noinertial}
\end{equation}

But why select this particular special case of the general $\mathcal{H}\mathcal{S}p(
2n) $ symmetry and give it the elevated status of a relativity group?\ \ 

Up to this point we have not made any comment on the particular
functional form of the Hamiltonian $H( p,q,t) $. The theorem is
silent on its form. Physical considerations lead to Hamiltonians
of many forms. For nonrelativistic electrodynamic, it is
\begin{equation}
H( p,q,t) =\frac{1}{2 m}{\left( p-\frac{\epsilon }{c}A( q,t) \right)
}^{2}+\epsilon  \phi ( q,t) 
\end{equation}

\noindent where in this equation $\phi ( q,t) $ is the electric
potential and $\epsilon $ is the charge.\ \ The canonical momentum
is related to the velocity through the expression
\begin{equation}
v( p,q,t) =\frac{p}{ m}-\frac{\epsilon }{m c}A( q,t) 
\end{equation}

\noindent and so the relationship between velocity and momentum
may be quite complex

For a broad class of problems in elementary classical mechanics,
the Hamiltonian is given simply by\ \ \ 
\begin{equation}
H( p,q,t)  =K( p) +V( q) =\frac{ p^{2}}{2 m}+V( q) .
\end{equation}

Hamilton's equations result in $v=\frac{ p}{ m} $\ \ and $\int v\cdot
d p=\frac{ p^{2}}{2 m}$ is the kinetic energy $K( p) $ and $-\int
f\cdot  d q=V( q) $\ \ is the potential energy.\ \ Energy is constant
in time as $\frac{\partial }{\partial  t}H( p,q) =0$.\ \ \ This
is but a most basic solution. An even more basic case is the inertial
state where\ \ $f=r=0$ and therefore $V( q) =0$. This state has
the property that, from (0),\ \ 
\begin{equation}
\tilde{H}( \tilde{p}) =H( p) + v\cdot p
\end{equation}

\noindent as both $v$ and $p$ are constant.\ \ Hamilton's equations
then transform as
\begin{equation}
\text{}\frac{d \tilde{q}( t) }{d t}=\frac{d q( t) }{d t}+v=\frac{\tilde{H}(
\tilde{p}) }{\partial \tilde{p}}=\frac{\partial H( p) }{\partial
p}+v,\text{}\frac{d \tilde{p}( t) }{d t}=\frac{d p( t) }{d t}=-\frac{\partial
\tilde{H}( \tilde{p}) }{\partial \tilde{q}}=0%
\label{mo: Hamilton's equations}
\end{equation}

\noindent and so the tilde equations are equivalent to the untilde'ed\ \ Hamilton
equations (0) with $\tilde{H}=H$ as functions .\ \ 

When the equations have this particularly simple form, extended
bodies that are constituted of multiple particles, such as a human
being, cannot\ \ distinguish between the moving and the rest frame
within the context of\ \ classical mechanics. This is important
as it allows us to travel on uniformly moving trains and jets. It
was for this reason that Galileo introduced this as a relativity
principle to explain why\ \ the earth could indeed by moving around
the sun while we have the Ptolemic perception that it is stationary.\ \ \ But
this is just a property of a very particular degenerate solution.
We know that such degenerate solutions break the symmetry of general
systems of equations. This leads to a strong relativity, $\tilde{H}=H$
and not the relativity or symmetry of the general set of equations
that has $\tilde{H}=H\circ \varrho ^{-1}$.\ \ Yet we have raised
these inertial states based on this property of a highly degenerate
specific solution to an almost exalted position in physics. An elementary
particle state simply does not distinguish between inertial and
noninertial states;\ \ it does not distinguish the inertial state
as having a very special status. It is just a degenerate solution.\ \ It
is the form of the equations, not a specific solution that must
be invariant under the group. 

Of course Galilean relativity is a limit of special relativity.
The Lorentz group contracts to the Euclidean group.\ \ Relativity
is fundamentally concerned with the concept of simultaneity and
the ordering of events by different observers in different physical
states. Special relativity has the property that simultaneity is
relative to the inertial state of observer state characterized by
$v$.\ \ It assumes, or rather, is silent about whether simultaneity
is affected by the relative noninertial state characterized by $f,r$.\ \ The
Minkowski metric\ \ 
\begin{equation}
d \tau ^{2}= d t^{2}-\frac{1}{c^{2}} d q^{2}.
\end{equation}

\noindent contracts to the degenerate Newtonian time line element
in the limit of small velocities relative to $c\text{}$. 
\begin{equation}
\gamma \mbox{}^{\circ}=\operatorname*{\lim }\limits_{c\rightarrow
\infty } d t^{2}( 1-\frac{v^{2}}{c^{2}} ) = d t^{2}.
\end{equation}

\noindent Simultaneity in the Galilean relativity limit is independent
of both the relative inertial and noninertial state and so we say
that it is absolute.\ \ 

General relativity locally has the same concept of simultaneity
as special relativity. It shows that gravity can be understood as
a curvature of a manifold with locally inertial frames, in which
special relativity continues to apply, and therefore simultaneity
depends only on the relative local inertial state.\ \ In a system
where there is only gravity, there are only locally inertial states;
all particles follow geodesics that are inertial trajectories in
the curved manifold and neighboring locally inertial frames are
related by the connection.\ \ The covariant derivative is relative
to these locally inertial frames related by the connection.\ \ General
relativity, like special relativity, is silent about simultaneity
and the clocks of particles in noninertial states due to other forces,
a simple example of which is an electron in a magnetic field. 

Just as Galilean relativity, that singles out inertial frames, is
the limit of special relativity, this simple theorem about Hamilton's
mechanics is the first pointer as the limit, to a relativity theory
in which simultaneity depends on the relative inertial and noninertial
state of the observer, characterized by the relative $v,f,r$ \cite{Low7},\cite{Low8}
This theory has a nondegenerate orthogonal Born metric \cite{born1},\cite{born2}
on extended phase space. This results in a relative simultaneity
between any states, inertial or noninertial.\ \ 

It may appear that a relativistic symmetry group on extended phase
space is not compatible with quantum mechanics. The quantum symmetry
is given by the projective representations that are equivalent to
equivalence classes of unitary representations of the central extension
of the group \cite{bargmann,mackey2}. 

Recall that the central extension of the inhomogeneous Euclidean
group, $\mathcal{I}\mathcal{E}( n) \simeq \mathcal{E}( n) \otimes
_{s}\mathcal{A}( n+1) $, is the Galilei group 
\[
\mathcal{G}a( n) =\overline{\mathcal{E}}( n) \otimes _{s}\mathcal{A}(
n+1) \otimes _{s}\mathcal{A}( 1) .
\]

\ \ The generator of the central $\mathcal{A}( 1) $ subgroup is
nonrelativistic mass that this group admits as an algebraic extension.\ \ \ The
central extension of the inhomogeneous Hamilton group $\mathcal{I}\mathcal{H}a(
n) =\mathcal{H}a( n) \otimes _{s}\mathcal{A}( 2n) $ is
\[
\check{\mathcal{I}\mathcal{H}a}( n) =\overline{\mathcal{H}a}( n)
\otimes _{s}\mathcal{H}( n+1) \otimes _{s}\mathcal{A}( 2) .
\]

The Galilei group is the inertial subgroup of this group with mass
one of the generators of the central $\mathcal{A}( 2) $ subgroup.\ \ The\ \ Weyl-Heisenberg
$\mathcal{H}( n+1) $ is parameterized by time, position, momentum
and energy and the Hermitian representation of its algebra are the
Heisenberg commutation relations.\ \ The projective representations
of the inhomogeneous Hamilton group are equivalence classes of the
unitary representations of this central extension.\ \ These may
be computed using the Mackey theorems for unitary representations
of semidirect product groups.\ \ One finds from this that the Hilbert
space is of the form $\text{\boldmath $\mathrm{H}$}\otimes {\text{\boldmath
$\mathrm{L}$}}^{2}( \mathbb{R}^{n+1},\mathbb{C}) $. Wave functions
are of the form $\psi ( q,t) $,or $\psi ( p,t) $ as we expect and
not wave functions of all the phase space degrees of $"\psi ( t,q,p,e)
"$.\ \ This is also the case in the relativistic generalization
\cite{Low5},\cite{Low6}.

The theorem that shows that\ \ Hamilton's equations have the symmetry
$\mathcal{S}p( 2n) \otimes _{s}\mathcal{H}( n) $ should not be surprising
as it is the intersection of a symplectic and affine symmetry, both
of which are fundamental in classical mechanics. This does not give
new results for classical mechanics but does give new insight into
noninertial frames. There is no reason to single out inertial frames
in Hamilton's mechanics as the equations are equally valid in inertial
and noninertial states provided the appropriate Hamilton function
is used. This does point to immediate relativistic \cite{Low5},
quantum \cite{Low8} and quantum relativistic theories \cite{Low6}
were the noniniertial symmetry in their context does have profound
implication.\ \ \ 

This paper is dedicated to Professor DeWitt-Morette for her lifelong
dedication to understanding the interplay between mathematics and
physics and giving an appreciation of that interplay to her students.
I would like to thank Peter Jarvis for discussions that have improved
the clarity of these ideas. 

\appendix\label{sp}

\end{document}